\documentstyle[twocolumn,aps,floats]{revtex}

\input psfig.sty

\begin{document}
\draft

\twocolumn[\hsize\textwidth\columnwidth\hsize\csname %
@twocolumnfalse\endcsname

\bibliographystyle{revtex}


\title{Linearity and Scaling of a Statistical Model for the Species
Abundance Distribution}

\author{Hector Garcia Martin and Nigel Goldenfeld}

\address{Department of Physics, University of Illinois at Urbana-Champaign \\1110 West Green Street, Urbana, IL, 61801}

\date{\today}


\maketitle

\begin{abstract}
We derive a {\it linear} recursion relation for the species
abundance distribution in a statistical model of ecology  and demonstrate the
existence of a scaling solution.
\end{abstract}
\vspace{0.2in}

]

\section{Introduction}

Understanding the relationship between  species richness in a biome
and its corresponding area is a long-standing problem in ecology, providing
important information about species richness, extinction of species due to
habitat loss and the design of reserves \cite{rosenzweig}.

Among the most usually cited mathematical functions relating the number
of different species (S) and the area they occupy (A) is the power law
form of the species area relationship (SAR): $S = c A^z$. In a paper by
Harte et al. \cite{harte1} this result was shown  to be equivalent to
assuming self-similarity in the distribution of species. Furthermore,
the species-abundance distribution, $P_0(n)$, the fraction of species with
$n$ individuals was found to satisfy a nonlinear recursion relation.

Banavar et al. went on to show that this model exhibits scaling data
collapse in the same way as observed in the two dimensional XY model
and in the power fluctuations in a closed turbulent flow
\cite{bramwell}, a result that follows from hyperscaling \cite{vivek1}.

The purpose of this paper is to show that the nonlinear recursion
relation can be recast as a {\it linear} recursion relation for the
species-abundance distribution that is much easier to handle; indeed,
since the equation governs a probability distribution, it natural to
expect that a linear equation is obeyed. By means of this recursion
relation we derive the scaling function {\it assumed} by Banavar et al.
\cite{banavar1}.

\section{The model and the nonlinear recursion relation}

\begin{figure}
\leavevmode\centering\psfig{file=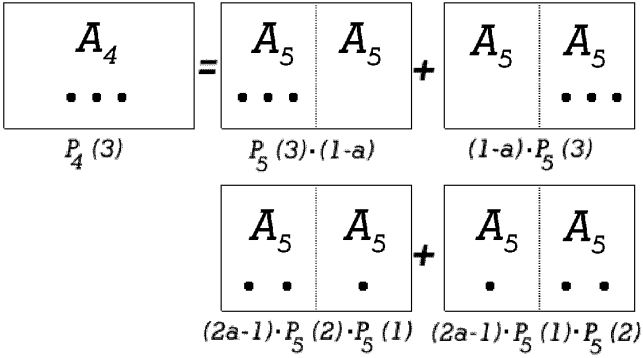,width=\columnwidth}
\caption[Harte]{ Explanation  of Equation~\ref{eq:nonlin} . Let's
consider the case $i = 4$ and $n = 3$. Circles correspond to individuals
 of a particular species found in a patch. On the left side there
 are three individuals in a patch $A_4$, on the right side all the
 possible ways in which those 3 individuals can be distributed in
the two patches $A_3$. The probability of finding three individuals in
a patch $A_4$ is then the addition of the probability that all the
individuals are on one side (prob. $1-a$) times the probability that
once all the individuals are on one side there are no individuals on
one side and there are three individuals on the other side
(prob. $1*P_5(3)$) plus the probability that the species are present on
both sides (prob. $2(1-a)$) times the probability that once the species
are present on both sides there are two individuals on one side
 and 1 individual on the other one
(prob. $P_5(2)*P_5(1)$). Taking $x=2(1-a)$ and $1-x=2a-1$ we find
$P_4(3)=xP_5(3)+(1-x)2P_5(2)P_5(1)$. This can be generalized to obtain
Equation~\ref{eq:nonlin}. Figure taken from ~\cite{harte1}.}
\label{fig-harte}
\end{figure}

In the model proposed by Harte et al. \cite{harte1} an area $A_0$ with a
number of species $S_0$ is considered. The number of individuals in
each species is described by $P_0(n)$, where  $S_0P_0(n)$ is the
expected number of species with n individuals. The area $A_0$ is
chosen to be in a shape of a rectangle with its length being $ \sqrt 2 $
times its width; such that by a bisection along the longer
dimension it can be divided in two rectangles of shape similar to
the original (see figure \ref{fig-harte}). $A_i = A_0/2^i$ is the area
of a rectangle after the {\it i}th bisection. If a species is present
in an area $A_i$, and nothing else is known about the species, there
are three possibilities: it might be present {\it only} on the right
subpartition of area $A_{i-1}$ (probability $P(R'|L)$), {\it only} on
the left one ($P(R|L')$) or in both ($P(R'|L')$). By symmetry
$P(R'|L)=P(R|L')$; and $a$ is defined as $P(R'|L) \equiv 1-a$.
The probability of finding a species on the right side, independently
of what happens on the left side is:
\begin{eqnarray}
P(R')&=&P(R'|L)+P(R'|L')=1-a+2a-1=a \\
&=&P(L') \textrm{ by symmetry} \nonumber
\end{eqnarray}
{\it Self similarity} is
introduced by stating that $a$ is independent of $i$, that is, scale.

Two conclusions can be
derived from this: a species area relationship of the kind $S = cA^z$
with $ a= 2^{-z}$  and  a recursion
relation for $P_i(n)$ (expected fraction of species with n
individuals for an area $A_i$, see figure \ref{fig-harte}) \cite{harte1}:
\begin{equation}
P_i(n) = xP_{i+1} + (1-x) \sum_{k=1}^{n-1}P_{i+1}(n-k)P_{i+1}(k)
\label{eq:nonlin}
\end{equation}
\noindent where $ x = 2(1-a) $. This recursion relation requires an
initial condition. It is supplied by defining a minimum patch $A_m =
A_0/2^m$, such that it contains on average only one individual (see
figure \ref{fig-boxes}). Consequently, $P_m(n)=\delta_{n1}$. This also
limits the maximum number of individuals that can be found in a patch
$A_i$ to $2^{m-i}$ so $P_i(n)=0$ for $n>2^{m-i}$.

\section{The linear relation}

\begin{figure}
\leavevmode\centering\psfig{file=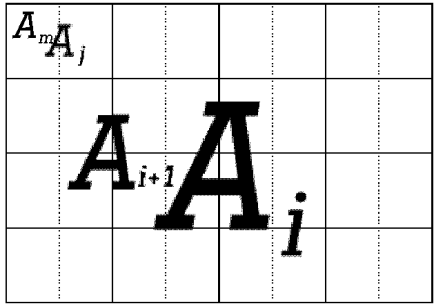,width=1.2\columnwidth}
\caption{$A_m$ is the minimum patch. $A_j$ in this case comprises two
minimum patches, but it can be of any size. In
Equation~\ref{eq:nonlin} the contributions to $P_i(n)$ come from
the two patches of size $A_{i+1}$, whereas in the case of the linear
recursion relation they come from the $2^{j-1}$ patches of size $A_{j}$.
}
\label{fig-boxes}
\end{figure}

Equation~\ref{eq:nonlin} is  nonlinear, and difficult to handle
analytically. The purpose of this section is to derive an equivalent
linear relation to calculate $P_i(n)$.  This derivation sums up
multiple patches at once, rather than proceeding strictly
hierarchically as in the original derivation.

We consider that the contributions to $P_i(n)$ come from several
($2^{j-i}$) patches of area  $A_j=A_i/2^j$ (``boxes'') instead of from
2 patches of area $A_{i+1}=A_i/2$ as before (see figure
\ref{fig-boxes}). The probability of finding $n$ individuals in $A_i$
is then the sum over the probabilities of finding $r$ of these
``boxes'' with the species present ($ R_j^i(r) $), multiplied by the
probability of finding $n$ individuals in these $r$ boxes ($ Q_j^i(r,n)
$):
\begin{equation}
P_i(n) = \sum_{r=1}^{2^{j-i}} R_j^i(r) Q_j^i(r,n)
\end{equation}
Note that the index $j$ is not summed over. It is arbitrary, indicating
the size of the ``box''. For $ j = i + 1 $ there are two boxes of area
$A_i/2$ and the original result of Harte et al is recovered, whereas
for $ j = m - 1 $ we will find a linear relation.  But before
establishing these results we calculate explicitly  $R_j^i(r)$ and
$Q_j^i(r,n)$:

\begin{itemize}
\item $Q_j^i(r,n)$ is the probability of finding $n$ individuals in $r$ boxes
of size $A_i/2^j$ in a total area $A_i$:
\begin{eqnarray}
\lefteqn{Q_j^i(r,n)=} & &                              \\ \nonumber
\sum_{n_1 \ldots n_n = 1}^{2^{m-j}}
  &(\prod_{l=1}^{r}P_j(n_l)) \delta(n-\sum n_k) &
  \hspace{5pt} r \le 2^{j-i}                            \\ \nonumber
  & 0 &  \hspace{5pt} r > 2^{j-i}
\end{eqnarray}
This formula is the probability of finding $n_1$ individuals in
the first box, $n_2$ in the second one, ... etc while the Kronecker
delta limits the possibilities to those that add up to the total number
of individuals $n$. $2^{j-i}$ is the maximum number of boxes and
$2^{m-j}$ is the maximum number of individuals in each box.

\item $R_j^i(r)$ is the probability of finding $r$ boxes of size $A_j$ in
which the species is present, in a total area $A_i$. This is just:
\begin{equation}
R^i_j(r) = P_{m+i-j}(r)
\end{equation}
This follows because the reasoning expressed in figure \ref{fig-harte} can be
applied to find the same recursion relation for  $R_j^i(r)$ as for
$P_i(n)$:
\begin{equation}
 R_j^i(r)=  xR_j^{i+1}(r) + (1-x) \sum_{k=1}^{r-1}R_j^{i+1}(k)R_j^{i+1}(r-k)
\end{equation}
The initial conditions do not change either, with $R_j^j(r) =
\delta_{r1}$. The only difference with the derivation for $P_i(n)$ is
that $r$ refers to the number of boxes (not individuals) and that the
recursion has to be applied $j-i$ times instead of $m-i$ times.

\end{itemize}

We can now check that for $j=i+1$ we find the same result as before:
\begin{eqnarray}
P_i(n) &=& \sum_r   R_1^i(r) Q_1^i(r,n) \\ \nonumber
       &=&   R_1^i(1) Q_1^i(1,n) +   R_1^i(2) Q_1^i(2,n)
\end{eqnarray}
Reading off from Equation (4):
\begin{eqnarray}
&Q_1^i(2,n) &= \sum_{k=1}^{n-1} P_{i+1}(k) P_{i+1}(n-k) \\
&Q_1^i(1,n) &= P_{i+1}(n) \\
&R_1^i(1)   &= x  \\
&R_1^i(2)   &= 1-x
\end{eqnarray}
Hence we obtain:
\begin{equation}
P_i(n) = x P_{i+1}(n) + (1-x) \sum_{k=1}^{n-1} P_{i+1}(k) P_{i+1}(n-k)
\end{equation}
as announced previously.
To obtain a linear relation we set $j=m-1$ and obtain:
\begin{equation}
Q_{m-1}(r,n)=\sum_{n_1,...n_r=1}^{2^{m-j}}
(\prod_{l=1}^{r}P_{m-1}(n_l)) \delta(n-\sum_i n_i)
\end{equation}
For $P_{m-1}(n)$ we only have the following possibilities:

\begin{equation}
P_{m-1}(n) =
\left\{\begin{array}{ll}
              x   &  n = 1 \\
              1-x &  n = 2 \\
          0   &  n \neq 1,2

\end{array}
\right.
\end{equation}

We find, denoting by $ q = n-r$  the number of boxes with two
individuals (factors of $P_{m-1}(2)$ in the equation above):
\begin{equation}
g(n,r)\equiv Q_{m-1}(r,n) = \frac{r!}{(r-q)!q!}x^{r-q}(1-x)^{q}
\end{equation}
The first factor is the number of possible configurations in which
there are $q$ boxes with two individuals and $n-q$ with one individual.
Finally we obtain:
\begin{equation}
P_i(n) = \sum_{r=1}^{2^{m-i-1}} P_{i+1}(r) g(n,r)
\end{equation}
which is a {\it linear} relation involving  $P_i(n)$ and $P_{i+1}(n)$.

\section{The scaling law}

Equation (13) allows us to derive the scaling law that
was assumed by Banavar et al. \cite{banavar1}:
\begin{equation}
P_i(n) = \frac{1}{n}f(\frac{n}{N_i^{\phi}})
\end{equation}
where $N_i$ $(= 2^{m-i})$ is the maximum number of individuals in an area
$A_i$ and $\phi = 1 - z$.

In order to achieve this, the following has to be done:
\begin{itemize}
\item First, find the {\bf continuum limit} for $g(r,n)$.
Since $g(r,n)$ is just a binomial distribution, it tends to a gaussian
for large $n$:
\begin{eqnarray}
\nonumber
g(n,r) =   \frac{r!}{(r-q)!q!}x^{r-q}(1-x)^{q} \\
 \approx   \frac{1}{\sqrt{2\pi r}} \frac{1}{\sqrt{x(1-x)}}
  \exp{\left(- \frac{1}{2}
\frac{(q-rx)^2}{rx(1-x)}\right)} \\ \nonumber
 = \frac{1}{\sqrt{\pi}\epsilon_{a,r} } \frac{1}{2{a}}
 \exp{\left(-\frac{(r-n/{2 a})^2}
 {\epsilon_{a,r}^2}\right)}\\  \nonumber
\end{eqnarray}
\begin{equation}
\epsilon_{a,r} = \sqrt{2(2a-1)(1-a) r/ (2a)^2}
\end{equation}
  $g(n,r)$ is the probability of finding $n$ individuals
in $r$ boxes. This probability is highly peaked
around $n=2ar$, since $2a$ ($= 1(1-a) + 1(1-a) + 2(2a - 1)$ )
 is the average of individuals per box. The more boxes
there are (bigger $r$) the sharper the peak. This means that
for large $r$ the only relevant values of $n$ are those near $n =
2{\bf a}r$ and  the expression given above for $g(n,r)$ is valid
for large $r$ (which implies large $n$).

\item  Second, rewrite everything in terms of a new variable $x$ and a
new probability density  $\overline{P}_i(x)$. $x$ replaces $n$ and is
the fraction of the total number of species: ${n}/{N_i}$ (which varies
from 0 to 1). $\overline{P}_i(n)$ is the density probability
${P_i(x)}/{(1/2^{m-i})}$, where $1/2^{m-i}$ is the distance between two
points in the new variable $x$. In this way all $P_i(n)$ can be
compared with each other in equal terms.
\end{itemize}

\begin{figure}
\leavevmode\centering\psfig{file=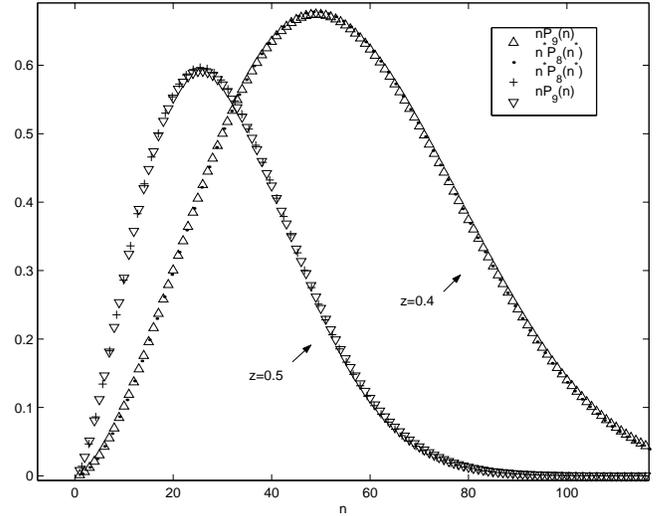,width=\columnwidth}
\caption{Scaling function $nP_0(n) = f(n/N_i^{\phi})$ for $m = 8$ and
$m=9$, and $z= 0.4$ and $z=0.5$. $n^* \equiv n/2^{\phi}=n/2a$. }
\label{fig-scaling}
\end{figure}

In terms of these new variables, the recursion relation can now be written as:
\begin{equation}
\overline{P}_i(x) = 2 \sum_{y=1/2^{m-i-1}}^{1} g(2^{m-i}x,2^{m-i-1}y) \overline
{P}_{i+1}(y)
\end{equation}
The continuum limit is found by taking $m$ (and consequently
the number of points $N_{i+1} = 2^{m-i-1}$) to an arbitrarily large
value and using the continuum limit of $g(r,n)$ as defined above. The
fact that the approximation for $g(r,n)$ is not a very good one for
small values of $n$ or $r$ is of little importance in the limit of large $m$ :
\begin{eqnarray}
\nonumber \overline{P}_i(x)&=&
 2 2^{m-i-1} \sum g(2^{m-i}x,2^{m-i-1}y)
\overline{P}_{i+1}(y) \underbrace{1/2^{m-i-1}}_{\Delta y} \\
&=& \int_{0}^{1} g^*(x,y) \overline{P}_{i+1}(y) dy
\end{eqnarray}
where $g^*(x,y) = 2 2^{m-i-1} g(2^{m-i}x,2^{m-i-1}y)$ and is equal to
$\frac{1}{a} \delta (y-x/a)$ in the limit of large $m$:
\begin{eqnarray}
g^*(x,y) &=& \frac{1}{\sqrt{\pi}} \frac{1}{2 { a} }
           \frac{1}{\epsilon_{y,a}}
           \frac{1}{\underbrace{2^{(m-i-1)/2}}_{\delta}}
       \exp[\frac{(y-x/a)^2}{\epsilon_y,a^2} \underbrace{2^{m-i-1}}_{\delta^2}]
\nonumber\\
         &=& \frac{1}{\sqrt{\pi}} \frac{1}{2 {a} }
           \frac{1}{\epsilon_{y,a} \delta}
        \exp{\frac{(y-x/a)^2}{(\epsilon_{y,a} \delta)^2} }
\end{eqnarray}
which is a standard representation of the Dirac delta function \cite{cohen} in the $x/a$
variable for $\epsilon_{y,a} \delta \rightarrow \infty $ (or $m
\rightarrow \infty $):
\begin{eqnarray}
 \lim_{m \rightarrow \infty} \int  g^*(x,y) f(x) dx =  f(ay) \nonumber
  \\
 \Rightarrow  \lim_{m \rightarrow \infty} g^*(x,y) = \frac{1}{a} \delta
(y-x/a)
\end{eqnarray}
This implies that
\begin{equation}
\overline{P}_i(x) = \frac{1}{a} \overline{P}_{i+1}(x/a)
\end{equation}
which is, in terms of $n$ and $P_i(n)$,
\begin{equation}
P_i(n) = \frac{1}{2a} P_{i+1}(n/2a)
\end{equation}
Since $ a = 2^{-z} $ and $ \phi = 1 - z $, multiplying the above
equation by $n$ and writing the explicit dependence of $P_i(n)$ on $N_i$
as $P_i(n)=P(n,N_i)$:
\begin{eqnarray}
&&nP(n,N_i)= \frac{n}{2a} P(n/2a,N_{i+1}) =\frac{n}{2a}
P(n/2a,N_i/2) \\ \nonumber
&& \Rightarrow f(n,N_i) \equiv nP(n,N_i) = \frac{n}{2^{\phi}}
P(n/2^{\phi},N_i/2)
\end{eqnarray}
which is by definition $f(n/2^{\phi},N_i/2)$.
Since $N_i$ is equal to a power of two this means that $nP{i}(n)$ is a
function only of $ n/{N_i^{\phi}} $:
\begin{equation}
P_i(n) = \frac{1}{n}f(\frac{n}{N_i^{\phi}})
\end{equation}
As can be appreciated  from the results above, a constant  $a$ (not
dependent on $i$) is necessary to obtain the scaling law: otherwise
$\phi$ would depend on $i$. In figure \ref{fig-scaling}  we exhibit the
scaling function for several $z$ and demonstrate the scaling law.
\acknowledgements
We thank John Harte and Annette Ostling for helpful discussions.  This
work was supported by the National Science Foundation through grant
NSF-DMR-99-70690.

\bibliography{Hector-paper1.1}

\begin{thebibliography}{1}

\bibitem{rosenzweig}
M. Rosenzweig, {\em Species Diversity in Space and Time} (Cambridge Univ.
  Press, Cambridge, 1995).

\bibitem{harte1}
J. Harte, A. Kinzig, and J. Green, Science {\bf {\bf 284}},  334  (1999).

\bibitem{bramwell}
S. Bramwell, P. Holdsworth, and J. Pinton, Nature {\bf {\bf 396}},  552
  (1999).

\bibitem{vivek1}
V. Aji and N. Goldenfeld, Phys. Rev. Lett. {\bf 86},  1007  (2001).

\bibitem{banavar1}
J. Banavar, J. Green, J. Harte, and A. Maritan, Phys. Rev. Lett. {\bf {\bf
  83}},  4112  (1999).

\bibitem{cohen}
C. Cohen-Tannoudji, B. Diu, and F. Laloe, {\em Quantum Mechanics} (John Wiley
  \& Sons, New York, 1977), Vol.~2, p.\ 1470.

\end{thebibliography}

\end{document}